\label{} 
\documentclass[preprint,showpacs,preprintnumbers,amsmath,amssymb]{revtex4}

\usepackage{amsmath,amssymb,amsfonts,times,graphicx,color}

\usepackage{graphicx}
\usepackage{dcolumn}
\usepackage{bm}

\def\barray{\begin{eqnarray}}
\def\earray{\end{eqnarray}}
\def\beq{\begin{equation}}
\def\eeq{\end{equation}}

\def\MA{\mathcal{A}}
\def\MC{\mathcal{C}}
\def\MO{\mathcal{O}}
\def\MI{\mathcal{I}}

\def\MN{\mathcal{N}}
\def\MM{\mathcal{M}}
\def\MW{\mathcal{W}}
\def\MT{\mathcal{T}}
\def\Area{{\rm Area}}

\begin{document}

\preprint{}

\title{An Operator Product Expansion for the Mutual Information in AdS/CFT} 

\author{Javier Molina-Vilaplana}
\affiliation{Department of Systems Engineering, Technical University of Cartagena, Dr Fleming S/N 30202 Cartagena, Spain}



\begin{abstract}
We investigate the behaviour of the mutual information $\MI_{AB}$ between two "small" and wide separated spherical regions $A$ and $B$ in the $\MN=4$ SYM gauge theory dual to Type IIB string theory in AdS$_5 \times S^5$.  To this end, the mutual information is recasted in terms of correlators of surface operators $\mathcal{W}\left( \Sigma\right) $ defined along a surface $\Sigma$ within the boundary gauge theory. This construction relies on the strong analogies between the twist field operators appearing in the replica trick method used for the computation of the entanglement entropy, and the disorder-like surface operators in gauge theories. In the AdS/CFT correspondence, a surface operator $\mathcal{W}\left( \Sigma\right) $ corresponds to having a D3-brane in AdS$_5 \times S^5$ ending on the boundary along the prescribed surface $\Sigma$. Then, a long distance expansion for $\MI_{AB}$ is provided. The coefficients of the expansion appear as a byproduct of the operator product expansion for the correlators of the operators $\MW(\Sigma)$  with the chiral primaries of the theory. We find that, while undergoing a phase transition at a critical distance, the holographic mutual information, instead of strictly vanishing, decays with a power law whose leading contributions of order $\MO(N^0)$, originate from the exchange of pairs of the lightest bulk particles between $A$ and $B$. These particles correspond to operators in the boundary field theory with the smallest scaling dimensions.
\end{abstract}


\maketitle

\section{Introduction}
Entanglement entropy and other related information-theoretic quantities such as mutual information, are by now regarded as valuable tools to study different phenomena in quantum field theories and many body systems \cite{holapp,holapp2}. These quantities provide a new kind of information that cannot be obtained from more standard observables such as two point correlation functions. Namely, both the entanglement entropy and the mutual information, are sensitive probes able to detect non-local signatures of the theory such as topological order which can not be detected by any local observable. Concretely, the mutual information $\MI_{AB}$ between two arbitrary regions $A$ and $B$ has certain advantages over the entanglement entropy. First, $\MI_{AB}$ can be viewed as an entropic correlator between $A$ and $B$ defined by,
\beq
\MI_{AB}=S_A + S_B - S_{A \cup B},
\label{mutinf}
\eeq
where $S_{A,\, B}$ is the entanglement entropy of the region $A (B)$ and 
$S_{A \cup B}$ is the entanglement entropy of the two regions. By its definition, $\MI_{AB}$ is finite 
and, contrarily to entanglement entropy, is non UV-cutoff dependent.
In addition, the subadditivity property of the entanglement entropy
states that when $A$ and $B$ are disconnected, then,
\beq\label{ssub}
S_A + S_B \geq S_{A \cup B},
\eeq
 which inmediatly leads to realize that $\MI_{AB} \geq 0$. Subadditivity is the most important inequality which entanglement entropy satisfies. A standard approach to compute 
 both the entanglement entropy and the mutual information makes use of the replica trick \cite{calcard,calcard2,calcard3}. Unfortunately, these calculations are notoriously difficult to carry out, even in the case of free field theories. 

In the context of the AdS/CFT \cite{adscftbib}-\cite{adscftbib4}, however, Ryu and Takayanagi (RT) have recently proposed a remarkably simple formula \cite{ryutak}-\cite{ryutak4} to obtain the entanglement entropy of an arbitrary region $A$ of a $d+1$ dimensional CFT which admits a classical gravity
dual given by an asymptotically AdS$_{d+2}$ spacetime. According to the RT formula, 
the entanglement entropy is obtained in terms of the area of a certain minimal surface  $\gamma_A$ in the 
dual higher dimensional gravitational geometry; as a result, the entanglement
entropy $S_A$ in a CFT$_{d+1}$ is given by the celebrated area law relation,
\beq\label{hologEE}
S_A = \frac{\Area(\gamma_A)}{4 G_N^{(d + 2)}},
\eeq
where $d$ is the number of space dimensions of the boundary CFT and
$\gamma_{A}$ is the $d$-dimensional static minimal
surface in AdS$_{d+2}$ such that $\partial A = \partial\, \gamma_A$. 
The  $G^{(d+2)}_{N}$ is the $d+2$ dimensional Newton constant. The RT formula 
provides a simple tool to prove the subadditivity of entanglement entropy from the 
properties of minimal surfaces \cite{headrick_sa}. Otherwise it has 
to be laboriously derived from the positive definiteness of the Hilbert space.

Here we consider the mutual information between two disconnected regions 
$A$ and $B$ in the ground state of an strongly coupled quantum field theory with
a gravity dual given by the AdS/CFT correspondence. Using (\ref{hologEE}) 
in (\ref{mutinf}), this quantity reads,
\beq\label{holoMI}
\MI_{AB}=\frac{1}{4 G_N^{(d + 2)}}\left[ \Area(\gamma_A) + \Area(\gamma_B)- \Area(\gamma_{A \cup B})\right],
\eeq
where $\Area(\gamma_{A \cup B})$ is the area of the minimal 
surface related to $A \cup B$. Recently, the holographic mutual information 
(\ref{holoMI}) has been considered in a quite remarkably amount of 
different settings \cite{headrick}-\cite{holMI7}.
A striking prediction for the holographic mutual information arises when analyzing the behaviour of 
the minimal surface $\gamma_{A \cup B}$. In \cite{headrick} it is shown how, 
for certain distances between the two regions, there are minimal surfaces $\gamma_{A \cup B}^{con}$
connecting $A$ and $B$. For those regimes, the holographic mutual information has a nonzero value 
proportional to the number of degrees of freedom in the gauge theory lying 
on the boundary of AdS$_{d+2}$.
However, when the separation between the 
two regions is large enough compared to their sizes, then a disconnected surface 
$\gamma_{A \cup B}^{dis}$ with,
\beq
\Area(\gamma_{A \cup B}^{dis}) = \Area(\gamma_A) + \Area(\gamma_B),
\eeq 
is both topologically allowed and minimal. In this case, (\ref{hologEE}) 
yields $S_{A \cup B} = S_A + S_B$ and a sharp vanishing of 
 $\MI_{AB}$ then occurs. This result is quite surprising from a quantum information 
point of view since, when the mutual information vanishes, the reduced density matrix 
$\rho_{A \cup B}$ factorizes into $\rho_{A \cup B} = \rho_A \otimes \rho_B$, 
implying that the two regions are completely decoupled
from each other and thus, all the correlations (both classical and quantum) 
between $A$ and $B$ should be rigorously zero. Indeed, 
it seems, at least counterintuitive, that all the correlations should strictly vanish 
at a critical distance, in a field theory in its large $N$ limit.
 This behaviour is a general prediction of the RT formula (\ref{hologEE})
which is valid for any two regions of any holographic theory. As a matter of fact, both 
(\ref{hologEE}) and (\ref{holoMI}) come from classical gravity in the bulk and 
provide the correct results to leading order in the $G_N$ expansion. 
When the boundary field theory is a large $N$ gauge theory,  
these terms are of order $N^2$. Thus, one might expect 
some corrections coming from quantum mechanical effects in the bulk theory, 
with the first correction appearing at order $N^0$ ($G_N^{\, 0}$) \cite{headrick}.
These $G_N^{\, 0}$ order corrections are small enough not to modify the shape of the 
surfaces and, as have been argued in \cite{maldacena}, jointly with the leading classical 
contributions, they obey the strong subadditivity condition.

In this note it is shown that, at least in the case that has been considered,
 the mutual information (\ref{holoMI}) between two disjoint regions $A$ and $B$ in the large separation regime, while undergoing a phase transition at a critical distance, instead of strictly vanishing, decays with a law whose leading contributions are given by the exchange of pairs of the lightest bulk particles between 
$A$ and $B$.  These bulk particles correspond to operators 
in the boundary field theory with small scaling dimensions as stated by the standard AdS/CFT 
dictionary \cite{adscftbib}-\cite{adscftbib4}. In order to achieve this result, first we propose to interpret the mutual information in terms of correlators of surface operators. These can be realized in terms of a probe D3-brane using the AdS/CFT correspondence \cite{surface}. An operator product expansion (OPE) for the long distance mutual information written in terms of these correlators is then provided. 

The expansion is in accordance with a recent proposal given in \cite{maldacena} where authors provide a long distance OPE for the mutual information $\MI_{AB}$ between disjoint regions inspired by an OPE for the mutual information in CFT previously discussed in \cite{headrick} and \cite{mi_ope3}. There, the expected leading contributions come from the exchange of pairs of operators $\MO_A, \MO_B$ located in $A$ and $B$ each with an small scaling dimension $\Delta$. The OPE reads as,
\beq
\label{eq0}
\MI_{AB}\sim \sum C_{\Delta}\left\langle \MO^{\Delta}_A\MO^{\Delta}_B \right\rangle ^2\sim \sum C_{\Delta} \left( \frac{1}{L}\right) ^{4\Delta}+\cdots,
\eeq
 where $L$ is the distance between $A$ and $B$ and $C_{\Delta}$ comes from squares of OPE coefficients. Thus, when considering a CFT theory with a gravity dual, one must deal with a quantum field theory in a fixed background geometry and the long distance expansion for the mutual information reduces to an expression similar to (\ref{eq0}), where now one should consider the exchange of the lightest bulk particles.  
 
The direct computation of the one-loop bulk corrections to the holographic entanglement entropy and R\'enyi entropies of two wide separated disjoint intervals in a 1+1 CFT has been explicitly addressed in \cite{hartnoll}. Here we ask if a simpler procedure can be used to learn, at least, some basic properties of the long range expansion of the $\MI_{AB}$ in higher dimensional theories.

\section{Mutual Information, Twist Operators and Surface Operators}
Our aim is to provide an OPE for the holographic mutual information in AdS$_5$ in terms of correlators of surface operators $\mathcal{W}\left(\Sigma\right) $ of the dual $\MN=4$ SYM gauge theory. To this end, in this section, we first present some general properties of $\MI_{AB}$ for subsystems that are weakly coupled to each other. We show a result that foreshadows the long distance expansion (\ref{eq0}) on very general grounds. Then we review the twist operators and their role in computing the entanglement entropy and the mutual information in quantum field theory through the replica trick method \cite{calcard,calcard2,calcard3}. Based on this, an OPE for the long distance mutual information is given. We also discuss on the strong analogies between the twist operators and the surface operators in gauge theories.

\subsection{Mutual Information between weakly coupled subsystems}
We assume, following \cite{headrick}, that the nearly factorized density matrix of two subsystems $A$ and $B$ separated by a distance $L$ much bigger than their characteristic sizes is given by, 
\beq
\label{eqa01}
\rho_{A \cup B} = \rho_{0} + \epsilon\, \rho_{1} + \epsilon^2\, \rho_{2},
\eeq
 where $\rho_{0}=\rho_{A}\otimes\rho_{B}$ with ${\rm tr}\, \rho_{0}=1$, ${\rm tr}\, \rho_{1}={\rm tr}\, \rho_{2}=0$ and $\epsilon \ll 1$. As a result, at order $\epsilon^2$, the entanglement entropy $S_{A \cup B}$ may be written as,
 \barray
 \label{eqa01b}
 S_{A \cup B} &=& - {\rm tr}\, [\rho_{A \cup B}\, \log\, \rho_{A \cup B}] \\ \nonumber
 &= &- {\rm tr}\, [(\rho_{0} + \epsilon\, \rho_{1}+ \epsilon^2\, \rho_{2})\, \log\, (\rho_{0} + \epsilon\, \rho_{1}+ \epsilon^2\, \rho_{2})]\\ \nonumber
 &\approx & - {\rm tr}\, (\rho_{0}\log \rho_{0}) - \epsilon\, {\rm tr}\, ((\rho_{1}+ \epsilon\, \rho_{2})\log\, \rho_{0}) - \epsilon^2\, {\rm tr}\, (\rho_{0}^{-1}\, \rho_{1}^{2})\\ \nonumber
 & = & S_A + S_B - \epsilon\, {\rm tr}\, ((\rho_{1}+ \epsilon\, \rho_{2})\log\, \rho_{0}) - \epsilon^2\, {\rm tr}\, (\rho_{0}^{-1}\, \rho_{1}^{2}),
 \earray 
 so, the mutual information at this order reads as,
 \beq
\label{eqa01b2}
\MI_{AB} \sim  \epsilon\, {\rm tr}\, ((\rho_{1}+ \epsilon\, \rho_{2})\log\, \rho_{0}) + \epsilon^2\, {\rm tr}(\rho_{0}^{-1}\, \rho_{1}^2).
\eeq
 
 Thus, it is straightforward to realize that at first order in $\epsilon$, the mutual information must vanish since $\epsilon$ could take either sign while $\MI_{AB}$ is always non-negative. Hence, the first non zero contribution to the mutual information is given by,
\beq
\label{eqa02}
\MI_{AB}\sim \epsilon^2\, {\rm tr}(\rho_{0}^{-1}\, \rho_{1}^2),
\eeq
which does not depend on $\rho_{2}$. It can be shown that the $\epsilon^2$ term in Eq.(\ref{eqa02})  does not generically vanish. Furthermore, since the non vanishing connected correlators between operators located in $A$ and $B$ are given by $\langle \MO_A(0)\, \MO_B(L)\rangle = {\rm tr}(\rho_1\, \MO_A\, \MO_B)$, then one might expect that,
\beq
\label{eqa03}
\MI_{AB}\sim \epsilon^2\, \langle \MO_A\, \MO_B\rangle^2 \sim   C\, \left( \frac{1}{L}\right)^{4\Delta},
\eeq
as far as $\langle \MO_A(0)\, \MO_B(L)\rangle \sim (1/L)^{2\Delta}$. This behaviour obeys the general bound given in \cite{wolf},
\beq
\label{eq_wolf}
\MI_{AB}\geq \frac{\langle \MO_A\MO_B\rangle^2}{2\Vert\MO_A\Vert^2\Vert\MO_B\Vert^2},
\eeq
 where $\Vert \MO \Vert$ is the absolute value of the maximum eigenvalue.

\subsection{Twist operators}
We consider now the computation of the entanglement entropy of a region (interval) $A$ in a (1+1)-dimensional CFT where $S_A$ is computed via the replica trick \cite{calcard,calcard2,calcard3} as,
\beq
\label{eq01}
S_A=-\partial_n\, {\rm tr}\, \rho_A^n\vert_{n=1} = -\partial_n\, \log\, {\rm tr}\, \rho_A^n\vert_{n=1},
\eeq
 with $\rho_A$ the reduced density matrix of the region $A$ and ${\rm tr}\, \rho_A = 1$. The method relies on the computation of ${\rm tr}\, \rho_A^n$ as a path integral over a $n$-sheeted Riemann surface, each sheet containing a copy of the CFT under consideration. This path integral happens to be equivalent to the path integral of the symmetric product of the $n$ copies of the original CFT (whose central charge is given by $nc$), defined on a single $\mathbb{R}^2$ sheet. Remarkably, ${\rm tr}\, \rho_A^n$ can be written as the two point function of two vertex-like point operators $\Phi_n^+(u)$ and $\Phi_n^-(v)$ called twist operators, inserted at the two boundary points ${u,v}$ of $A$ in the path integral, i.e, 
\beq
\label{eq02}
{\rm tr} \rho_A^n = \left\langle \Phi_n^+(u)\, \Phi_n^-(v)\right\rangle. 
\eeq

The twist operators are actually primary operators with scaling dimensions $\Delta_n = \frac{c}{12}(n-\frac{1}{n})$ related to the central charge of the CFT and the number of replicas $n$. They account for the conical singularities appearing as one joins the $n$ copies of the CFT in the $n$-sheeted surface formulation of the path integral.

 In $d+1$ dimensions, one may also compute ${\rm tr}\, \rho_A^n$ as a path integral over an $n$-sheeted Riemann surface. This multi-sheeted surface has a conical singularity along the boundary $\partial A$ of the region $A$ for which one is computing the entropy. It is expected that this path integral can be written as a path integral on a single-sheeted surface with an inserted twist-like operator $\MT_n\left[\partial A\right]$ defined along the boundary $\partial A$. Thus, ${\rm tr} \rho_A^n = \left\langle \MT_n\left[\partial A\right] \right\rangle $ and, in absence of further operator insertions, ${\rm tr} \rho_A = 1$. Here, the operator $ \MT_n\left[\partial A\right]$ is no longer point-like, becoming instead an extended operator such as a line operator in 2+1 dimensions or a surface operator in 3+1 dimensions.
 
 As pointed out in \cite{swingle10}, a key realization about twist fields in a (1+1) dimensional CFT is their resemblance with operators builded as the exponential of a massles field, i.e, a vertex operator in a free boson CFT. In practice, the construction and properties of these twist fields beyond (1+1) dimensions is poorly understood\footnote{In higher dimensions, the replica trick provides only a formal definition of the the twist operators and much of their properties are unknown. See \cite{myers} for very recent advances on these topics.}. Nevertheless, let us briefly discuss on how these higher dimensional $\MT_n\left[\partial A\right]$ operators exhibit significant analogies with extended operators in gauge theories.
 
Assuming a vertex-like functional structure for a higher dimensional twist-field amounts to argue that it is the exponential of a certain type of massles spatial $(d-1)$-form $F^{(d-1)}$,
\beq
\label{eq03}
\MT_n\left[\partial A\right]=\exp\left(i \alpha_n\, \int_{\partial A} F^{(d-1)}\right),
\eeq  
 where $\alpha_n$ must be fixed so as to obtain the correct prefactor for the entanglement entropy, which in a strongly coupled field theory is proportional to $N^{\, 2}$. As long as the region $A$ is compact, it is easy to show that $F^{(d-1)}$ and thus $\MT_n\left[\partial A\right]$ has a "gauge symmetry" \cite{swingle10},
\beq
\label{eq04}
 F^{(d-1)}\to  F^{(d-1)} + d\Lambda^{(d-2)},
\eeq  
 with $\Lambda^{(d-2)}$ an arbitrary spatial $(d-2)$-form. Let us to further illustrate the \emph{ansatz} in Eq.(\ref{eq03}) by considering a scalar field theory $\phi$ in 3+1 dimensions and a set of $n$ replica fields $\lbrace \phi_n \rbrace$. These fields amount to a representation of the cyclic permutation subgroup of $\mathbb{Z}_n$ generated by the twist operator $\MT_n\left[\partial A\right]$,
 \beq
 \MT_n\left[\partial A\right]: \phi_n \longrightarrow \phi_{n \pm 1} \quad {\rm mod}\, n.
 \eeq

In other words, the twist operator $\MT_n\left[\partial A\right]$ is the analog in the original multi-sheeted surface of moving from one sheet to the next (previous) one. Now, it is useful to introduce the linear combination of the replica fields,
\beq
\widetilde{\phi}_k \equiv \sum_{j=1}^{n}\, e^{2\pi i\frac{k}{n} j} \, \phi_j, \quad k=0,1,\cdots, n-1,
\eeq
which are phase shifted by the factor $\lambda_k= e^{2\pi i k/n}$ as they encircle the codimension-2 spacetime region on which the twist operator is defined, i.e, they diagonalize the twist operator,
\beq
\MT_n\left[\partial A\right]\, \widetilde{\phi}_k = \lambda_k\, \widetilde{\phi}_k.
\eeq

Namely, the twist operator $\MT_n[\partial A]$ can be written as a product of operators $\MT_{n,k}[\partial A]$ acting only on $\widetilde{\phi}_k$,
\beq
\MT_n\left[\partial A\right]= \prod_{k=0}^{n-1}\, \MT_{n,k}[\partial A],
\eeq
with $\MT_{n,k}[\partial A] \widetilde{\phi}_{k'}=\widetilde{\phi}_k $ if $k \neq k'$ and $\MT_{n,k}[\partial A] \widetilde{\phi}_{k}= \lambda_k\, \widetilde{\phi}_k$. 

The way the field $\widetilde{\phi}_k $ picks up the phase shift $\lambda_k$ resembles the Aharonov-Bohm effect. Namely, since $\vert \lambda_k \vert=1$, one might introduce 2-form gauge fields $F^{(k)}$ to give account for these phase shifts. These fields are normal gauge fields with a singular behaviour along the codimension-2 locus where the twist operator is defined. In a (3+1) dimensional theory, this locus amounts to a closed 2-dimensional surface. Therefore, the twist operator $\MT_n\left[\partial A\right]$ would be some 2-dimensional \emph{surface} operator introducing a branch cut in the path integral over the $n$-fold replicated theory. In case that the entangling surface $\partial A$ is a static $S^2$ sphere, the twist operator residing on it, acts by opening a branch cut over the ball on the interior.

Noticing that $\mathbb{Z}_n$ acts on $\lbrace \widetilde{\phi}_k \rbrace$ as a global $U(1)$ charge symmetry,  the twist operator $\MT_{n,k}[\partial A] $ can be defined by (see  Eq.(\ref{eq03})),
\beq
\MT_{n,k}\left[\partial A\right]\sim \exp\left(i \int_{\partial A}\, F^{(k)} \right),
\eeq 
where $F^{(k)}$ encodes the flux which generates the phase shift $\lambda_k$. A similar analysis has been carried out in \cite{casini0} in the two dimensional case, when the twist field is point-like and local. There authors first discussed the interpretation of the twist fields as vortex-like operators. 
  
 To finalize, we also note that the mutual information between two regions $A$ and $B$ can be written in terms of the twist operators $\MT_n\left[\partial A\right]$ and $\MT_n\left[\partial B\right]$ as, 
\beq
\label{eq06}
\MI_{AB}=\partial_n\left[\log\, \frac{ \left\langle \MT_n\left[\partial A\right]\,  \MT_n\left[\partial B\right] \right\rangle }{\left\langle \MT_n\left[\partial A\right] \right\rangle \left\langle \MT_n\left[\partial B\right] \right\rangle} \right]_{n=1},
\eeq
which amounts to compute the connected correlation function between $\MT_n\left[\partial A\right]$ and $\MT_n\left[\partial B\right]$. As an example, in CFT$_2$, if one considers two disconnected intervals $A=[u_1,\, v_1]$, $B=[u_2,\, v_2]$ $(u_1 < v_1< u_2 < v_2)$ such that $\partial A=\lbrace u_1,\, v_1\rbrace$ and $\partial B=\lbrace u_2,\, v_2\rbrace$, then Eq.(\ref{eq06}) may be written as \cite{headrick},
\beq
\label{eq07}
\MI_{AB}=\partial_n\left[\frac{1}{n-1}\log\, \frac{ \left\langle \Phi^{+}_n(u_1)\, \Phi^{-}_n(v_1)\, \Phi^{+}_n(u_2)\, \Phi^{-}_n(v_2)\,\right\rangle }{\left\langle  \Phi^{+}_n(u_1)\, \Phi^{-}_n(v_1) \right\rangle \left\langle  \Phi^{+}_n(u_2)\, \Phi^{-}_n(v_2) \right\rangle} \right]_{n=1},
\eeq
where $ \Phi^{+}(u),\, \Phi^{-}(v)$ are the point-like twist operators mentioned above.

\subsection{Long distance expansion for the Mutual Information}
\label{LongDistanceMI}
 It has been argued in \cite{headrick} that the minimal area prescription in Eq.(\ref{hologEE}) and Eq.(\ref{holoMI}), though providing tempting hints about the structure of correlations in holographic theories at order $G_N$, hides an important part of that structure in  situations such as the long distance regime of the mutual information. Here we argue that it might result helpful to rephrase these quantities in terms of correlators of twist operators (Eq.(\ref{eq06})) since, once taken this approach, it is in principle possible, to have an OPE of these correlators from which $(G_N)^q$, $q\geq0$ corrections to Eq.(\ref{holoMI}) might be obtained. Let us settle on this claim. The twist operator $\MT_n\left[\partial A\right]$ can be expanded in a series of local operators $\MO^{A}_i$ when probed from a distance $L$ much larger than the characteristic size $a$ of the region $A$ as,
 \beq
 \label{eqa04}
 \MT_n\left[\partial A\right] = \langle \MT_n\left[\partial A\right] \rangle\, \left( 1 + \sum_i\, \MC_i^{A}(a, \Delta_{i},0)\, \MO^{A}_i(0)\right), 
 \eeq 
where $\Delta_{i}$ are the conformal dimensions of the operators. The exact form of the expansion coefficients $\MC_i^{A}(a, \Delta_{i},0)$ is unknown but generally, they should depend both on the scale $a$ and of the reference point at which  the operator $\MO^{A}_i$ is inserted. Here, we have choosen the reference point as the center of the spherical region enclosed by the twist operator, i.e, the origin. For the sake of subsequent arguments in this paper, the operators $\MO^{A}_i$ are conformal primaries inserted at a single copy of the $n$-folded replica trick construction, while in general, they consist in products of two or more of such operators inserted at the same point but in different copies of the CFT \cite{myers}.

 A similar expansion also holds for the twist operator $\MT_n\left[\partial B\right]$ defined along the boundary of a region $B$ with characteristic size $a$ located at a distance $L$ from the origin,
\beq
 \label{eqa05}
 \MT_n\left[\partial B\right]= \langle \MT_n\left[\partial B\right] \rangle\, \left( 1 + \sum_j\, \MC_j^{B}(a,\Delta_j,L)\, \MO^{B}_j(L)\right).
 \eeq 
 
Thus, the OPEs and their coefficients $ \MC_i^{A},\,  \MC_j^{B}$ appear as one replaces the regions $A$, $B$ by a sum of local CFT operators\footnote{Henceforth, we simplify the notation by omitting the explicit dependence on the scale $a$, the conformal dimensions and insertion points of the expansion coefficients $ \MC_i^{A}$, and $\MC_j^{B}$.}. Assuming that the vacuum expectation value of a single operator $\langle \MO \rangle = 0$, the connected correlator in Eq.(\ref{eq06}) can be written as,
\beq
 \label{eqa06}
 \log \frac{\langle \MT_n\left[\partial A\right]\, \MT_n\left[\partial B\right]\rangle}{\langle \MT_n\left[\partial A\right]\rangle \langle \MT_n\left[\partial B\right]\rangle} \sim  \sum_{i,j}\, \MC_i^{A}\, \MC_j^{B}\, \langle \MO^{A}_i(0)\, \MO^{B}_j(L)\rangle .
 \eeq 
 
However, recalling Eqs.(\ref{eqa02})-(\ref{eqa03}), one notices that this OPE for the mutual information should not be valid, as only involves ${\rm tr}(\rho_1)$ terms $(\sim \langle \MO\, \MO \rangle)$ contrarily to the expected ${\rm tr} (\rho_1^2)$ ones $(\sim \langle \MO\, \MO \rangle^2)$. Let us  fix this  point by focusing on the 1+1 CFT case. If one performs a sort of OPE such as the one given by Eq.(\ref{eqa06}) on the quantity within the brackets in Eq.(\ref{eq07}), then the computation of $\MI_{AB}$ singles out the term that is linear in $(n-1)$. It turns out that terms $\langle \MO\, \MO \rangle$ in that expansion are proportional to $(n-1)^2$ as shown in \cite{headrick}, and therefore, their contribution vanishes after doing the derivative and taking the $n \to 1$ limit\footnote{I thank Juan M. Maldacena for some clarifications on this subject}. As a result, one might be compelled to consider an alternative OPE for $\MI_{AB}$ which, while using the long distance expansion for correlators of twist fields, takes into account  Eqs.(\ref{eqa02})-(\ref{eqa03}). 
 
We first  notice that the long distance expansion for the operator $ \MT_n\left[\partial A\right]$ with a chiral primary operator (CPO) $\MO^{B}_{k}$ inserted at $\partial B$ is given by,  
 \beq
 \label{eqa07}
 \frac{\langle \MT_n\left[\partial A\right]\, \MO^{B}_{k}(L)\rangle}{\langle \MT_n\left[\partial A\right]\rangle} = \MC_k^{A}\,  \langle \MO^{A}_{k}(0)\, \MO^{B}_{k}(L)\rangle \sim \MC_k^{A}\, \left( \frac{1}{L} \right)^{2\Delta_k},
 \eeq  
where $L$ is the distance between regions $A$ and $B$ and $\Delta_k$ is the scaling dimension of the CPO $\MO^{B}_{k}$. Similarly, the long distance expansion for the correlator of $\MT_n\left[\partial B\right]$ with a CPO $\MO^{A}_{m}$ inserted at $\partial A$ is given by,
\beq
 \label{eqa08}
 \frac{\langle \MO^{A}_{m}(0)\,  \MT_n\left[\partial B\right]\rangle}{\langle \MT_n\left[\partial B\right]\rangle} = \MC_m^{B}\,  \langle \MO^{A}_{m}(0)\, \MO^{B}_{m}(L)\rangle \sim \MC_m^{B}\, \left( \frac{1}{L} \right)^{2\Delta_m},
 \eeq  
with $\Delta_m$ the scaling dimension of the CPO $\MO^{A}_{m}$. As a consequence, it results reasonable to propose a  long distance OPE for the mutual information which jointly takes into account the long distance correlators of each one of the twist fields with all the CPO which one might find inserted on the other region. This can be written as,
\barray
 \label{eqa09}
\MI_{AB} &\sim & \partial_n\, \left[ \sum_{k,\, m}\, \frac{\langle \MT_n\left[\partial A\right]\, \MO^{B}_{k}(L)\rangle}{\langle \MT_n\left[\partial A\right]\rangle}\, \frac{\langle \MO^{A}_{m}(0)\,  \MT_n\left[\partial B\right]\rangle}{\langle \MT_n\left[\partial B\right]\rangle}\right]_{n=1}   \\ \nonumber
 &= & \sum_{k}\, C_{k}\, \left( \frac{1}{L^{2}}\right)^{2\Delta_k}  + \sum_{k \neq m}\, \partial_n\left[ \MC_k^{A}\, \MC_m^{B}\right]_{n=1} \, \left( \frac{1}{L^{2}}\right)^{\Delta_k + \Delta_m},
 \earray 
 with $C_k = \partial_n\left[ \MC_k^{A}\, \MC_k^{B}\right]_{n=1} $. This "OPE" accomodates to the very general requeriments for the behaviour of $\MI_{AB}$ between weakly coupled regions showed above, while its coefficients are a byproduct of the OPE between the twist fields and the CPO of the CFT.

At this point it is worth to note that, while little is known about twist fields in higher dimensional CFTs, not to say about the coefficients $\MC_k$ of the OPE. As discussed above, those seem to be line or surface-like operators of a sort with analogous properties to the better known line and surface operators of gauge theories. Therefore, it might result tempting to access the properties of the mutual information in higher dimensional theories through the properties of these higher dimensional gauge operators, especially in situations where the benefits of computing through the AdS/CFT correspondence are manifest. This also relates to the question of, up to what extent, some information theoretic quantities such as the mutual information might determine the underlying QFT \cite{casini}. In this sense, one may realize following \cite{casini}, that as the entropy $S_{A \cup B}$ for very distant regions $A$ and $B$ approaches the sum of entropies $S_A + S_B$, the vacuum expectation value (VEV) of product of operators $\mathcal{W}_A$ and $\mathcal{W}_B$ defined on $A$ and $B$, factorizes into the product of VEV, so the exponential ansatz for $\MI_{AB}$,
\beq
e^{\mu\, \MI_{AB}} = \frac{\langle \mathcal{W}_A\, \mathcal{W}_B \rangle}{\langle \mathcal{W}_A\rangle\, \langle \mathcal{W}_B \rangle},
\eeq
where $\mu$ is a number, is exactly what one might expect in order to account for the clustering properties of correlators and entropies. This ansatz 
is a mapping that must respect both Poincar\'e symmetry and causality. The causality constraint imposes that $\mathcal{W}_A$, which in principle is a product of operators fully supported on $A$, should be the same for all the spatial surfaces with the same boundary as $\partial A$. This implies that $\mathcal{W}_A$ must be localized on $\partial A$, which in more than one spatial dimensions, once more suggests that it may be some kind of generalized 'Wilson loop' operator of the theory under consideration.

Here, it is worth to recall that in Eqs.(\ref{eqa07},\, \ref{eqa08},\, \ref{eqa09}), one must deal with the correlators of the twist operators $\MT_n$ with the primary operators $\MO$ of the theory inserted at a single copy of the CFT, for instance, the first of the $n$ copies. At this point, we follow \cite{myers} in order to construct (at least formally) a surface-like effective twist operator $\widetilde{\MT}_n$ which only acts within the first copy of the CFT by reproducing any correlator of the form,
\beq
\langle \MT_n\,  \MO\rangle = \langle \widetilde{\MT}_n\,  \MO\rangle_{\mathbf{1}}, 
\label{effectwist}
\eeq
where the subscript on the second correlator means that its computation is carried out on the first single copy of the CFT. As in the two dimensional case, it is reasonable to assume that some roles of these effective twist operators, such as to impose the correct boundary conditions on the fields of the theory through their vortex-like singularities, are common to codimension-2 surface-like operators of the CFT. Under this assumption, our approach here will consist in modifying Eq.(\ref{eqa09}) by means of the effective twist operator construction in Eq.(\ref{effectwist}) and then to supersede $\langle \widetilde{\MT}_n\left[\Sigma \right] \,  \MO\rangle_{\mathbf{1}}$ with the correlation function $\langle \MW\left[\Sigma \right] \,  \MO\rangle$ between a surface operator $\MW\left[\Sigma \right]$ of the CFT and a primary operator $\MO$, with $\Sigma$ as the spatial surface on which the operators are defined.

As a consequence, provided they can be computed, one may probe the long distance behaviour of $\MI_{AB}$ by means of the OPE between the surface operators $\MW\left[ \partial A,\, 0\right]$, $\MW\left[ \partial B,\, L\right]$ and the CPO of the gauge theory under consideration, 
\barray
\frac{\langle \mathcal{W}\left[\partial A, 0\right]\, \MO^{B}_{k}(L)\rangle}{\langle \mathcal{W}\left[\partial A, 0\right]\rangle} & = & \widetilde{\MC}_k^{A}\,  \langle \MO^{A}_{k}(0)\, \MO^{B}_{k}(L)\rangle \sim \widetilde{\MC}_k^{A}\, \left( \frac{1}{L} \right)^{2\Delta_k},\\ \nonumber
\frac{\langle \MO^{A}_{m}(0)\,  \mathcal{W}\left[\partial B, L\right]\rangle}{\langle \mathcal{W}\left[\partial B, L\right]\rangle} & = & \widetilde{\MC}_m^{B}\,  \langle \MO^{A}_{m}(0)\, \MO^{B}_{m}(L)\rangle \sim \widetilde{\MC}_m^{B}\, \left( \frac{1}{L} \right)^{2\Delta_m},
\earray
where coefficients $\widetilde{\MC}_i^{A},\, \widetilde{\MC}_j^{B}$ depend explicitly on the characteristic size of the spatial regions $A$ and $B$ and both the insertion points and the scaling dimensions of the CPO. Finally, the long distance expansion for $\MI_{AB}$  written in terms of these correlators  reads as,
\barray
\label{su_mi:ope}
\MI_{AB} &\sim &  \sum_{k, m}\, \frac{\langle \mathcal{W}\left[\partial A,0\right]\, \MO^{B}_{k}(L)\rangle}{\langle \mathcal{W}\left[\partial A, 0\right]\rangle}\, \frac{\langle \MO^{A}_{m}(0)\,  \mathcal{W}\left[\partial B, L\right]\rangle}{\langle \mathcal{W}\left[\partial B, L\right]\rangle} \\ \nonumber
&=& \left[ \sum_{k}\, \widetilde{C}_{k}\, \left( \frac{1}{L^{2}}\right)^{2\Delta_k}  + \sum_{k \neq m}\, \widetilde{\MC}_k^{A}\, \widetilde{\MC}_m^{B}\, \left( \frac{1}{L^{2}}\right)^{\Delta_k + \Delta_m}\right],
\earray
where $\widetilde{C}_k =  \widetilde{\MC}_k^{A}\, \widetilde{\MC}_k^{B}$. The sums arise by considering all the possible local primary operators of the CFT which one might expect to find inserted at each one of the surfaces $\partial A, \partial B$. This is precisely the scenario that will be considered in the remainder of this paper. As in the two dimensional case \cite{mi_ope2}, the leading contributions to $\MI_{AB}$ in Eq.(\ref{su_mi:ope}) are controlled by the conformal primaries of the theory. Nevertheless, while in (1+1) CFT the expansion coefficients only depend on the correlation function of these operators, in the higher dimensional case, these coefficients non trivially depend on the geometry of the regions $A$ and $B$ as has been mentioned above. 

\section{Mutual Information in $\MN=4$ SYM from AdS$_5 \times$ S$^5$}
We analyze the the mutual information between two static spherical 3-dimensional regions $A$ and $B$ with radius $a$ and separated by a distance $L\gg a$, in the $\MN = 4$ SYM theory dual to Type IIB superstring theory on AdS$_5 \times S^5$. To this aim, we first briefly review the holographic realization of surface operators in the gauge theory and then, using the arguments exposed above, a long distance expansion of the mutual information in terms of the correlators between these operators and the chiral primaries of the theory is provided.

\subsection{Surface Operators in $\MN=4$ SYM gauge theory.}
There are different kinds of operators in a 4-dimensional gauge theory attending to the spacetime locus on which they are supported. Codimension-4 operators are point-like local operators that have been extensively studied in the AdS/CFT correspondence. Codimension-3 operators are one dimensional operators such as the Wilson and t'Hooft loops. Two dimensional surface operators $\MW(\Sigma)$ are defined along a codimension-2 surface $\Sigma \subset \MM$, where $\MM$ is the spacetime manifold on which the theory is defined \footnote{For previous work involving codimension two singularities in a gauge theory, see \cite{preskill}}. The later were studied by Gukov and Witten in the context of the geometric Langlands program, where they classified them in order to understand the action of S-duality \cite{gukov_witten1, gukov_witten2}.

In a theory with a gauge group $G=U(1)$ \footnote{For simplicity we have considered a $U(1)$ gauge field, but indeed, for $U(N)$, there are different types of surface operators labelled by partitions of $N$.}, surface operators are disorder operators which, like t'Hooft operators, can be defined by requiring the gauge field to have a prescribed vortex-like singularity along the surface $\Sigma$:
\beq
F = 2\pi \alpha\, \delta_{\Sigma} + {\rm smooth},
\eeq
where $F$ is the gauge field curvature 2-form and $\delta_{\Sigma}$ is 2-form delta function that is Poincar\'e dual to $\Sigma$. 
Then, the new path integral is over fields with this prescribed singularity along $\Sigma$. This amounts to introduce a phase factor $\eta$ in the path integral by inserting the operator,

\beq
\exp\left(i\, \eta\, \int_{\Sigma}\, F \right). 
\eeq

Thus, one needs to consider the path integral with a special prescribed singularity along a codimension-2 manifold $\Sigma$.  The fields of the theory acquire the phase factors $\eta$ as they encircle the codimension-2 surface $\Sigma$ due to their singular behaviour near it. As puzzling as they may seem, these singularities are rather ubiquitous in theories with vortex-like disorder operators such as the discontinuities induced on the fields of the theory by twist (or effective twist) operators in a higher dimensional CFT.  As in the case of a two dimensional CFT, these discontinuities are consistent as far as the correlation functions of physical operators remain well defined. 

Some remarkable calculations involving disorder-like surface operators in the context of the AdS/CFT correspondence have been carried out both in a four dimensional gauge theory \cite{surface} and in a three dimensional theory \cite{vortex}. In the large $N$ and large t'Hooft coupling $\lambda$ limit of the four dimensional $\MN=4$ SYM theory, the vortex-like surface operators can be holographically described in terms of a D3-brane in AdS$_5 \times S^5$ with a worldvolume $Q\times S^1$, where $S^1 \subset S^5$ and $Q \subset$ AdS$_5$ is a volume minimizing 3-manifold with boundary,
\beq
\partial Q =\Sigma \subset \MM.
\eeq
Likewise, the holographic M-theory representation of a one dimensional vortex-like operator in the ABJM three dimensional $\MN=6$ supersymmetric Chern-Simons theory \cite{abjm}, amounts to an M2-brane ending along one dimensional curve on the boundary of AdS$_4 \times S^7/\mathbb{Z}_k$. 

Both descriptions are a probe brane approximation. Those are valid when the vortex-like operators under consideration have singular values only in the $U(1)$ factor of the unbroken gauge group $U(1) \times SU(N-1)$, which is the case that will be considered in this paper. When the singular behaviour of the gauge fields are not such specifically restricted, then the disorder operators correspond to arrays of branes from which a pure geometric description in terms of regular "bubbling" geometries can be obtained \cite{gomis2}.

\subsection{Long distance expansion for the Holographic Mutual Information}
 We go back to the arguments given at the end of Section 2 and thus consider the OPE for the mutual information (\ref{su_mi:ope}) written in terms of the correlators of surface operators $\mathcal{W}(\Sigma)$ with the chiral primary operators $\MO_{k}$,
\beq
\label{eq9}
 \frac{\left\langle \mathcal{W}(\Sigma,0)\, \MO_{k}(L)\right\rangle }{\left\langle \MW(\Sigma, 0) \right\rangle }, 
\eeq   
where $\Sigma = \partial A$ or $\partial B$, are two static 2-dimensional spherical regions with radius $a$, $\Delta_{k}$ is the scaling dimension of the primary operator and $L$ is distance between them. 
 
 As stated above, in the supergravity approximation, when $N\gg 1$ and the t'Hooft coupling $\lambda \gg 1$, the surface operator $\mathcal{W}(\Sigma)$ is related with a D3-brane $\subset$ AdS$_5$ ending on the boundary of the spacetime with a tension given by $T_{D3}=N/2\pi^2$ (in the units where the AdS$_5$ radius $R^4=1$).
 
 The correlator (\ref{eq9}) is calculated by treating the brane as an external source for a number of propagating bulk fields in AdS and then computing the brane effective action $S_{D3}$ for the emission of  the supergravity state associated to the operator $\MO_k$ onto the point on the boundary where it is inserted \cite{surface}\footnote{See also \cite{malda98}.}. The prescription to compute this correlator is to functionally differentiate $S_{D3}$ with respect to the bulk field $s_k$. This yields a correlator which scales with the distance $L$ as,
 \beq
 \label{eq9bis}
   \frac{\left\langle \mathcal{W}(\Sigma,0)\, \MO_{k}(L)\right\rangle }{\left\langle \mathcal{W}(\Sigma, 0) \right\rangle } = -\left. \frac{\delta S_{D3}}{\delta s_k}\right|_{s_k=0}=\widetilde{\MC}_{k}\, \left( \frac{1}{L}\right)^{2\Delta_k}.
 \eeq
 
 Thus, in the following, the quantities that one might be concerned to compute, are the OPE coefficients $\widetilde{\MC}_k$. We will outline the calculations just below, but full details of it, can be found in \cite{surface}.  As a result, our proposal for the long distance expansion of the mutual information given in Eq.(\ref{su_mi:ope}), may be holographically realized in terms of the mutual exchange of bulk particles between the codimension-2 regions $\partial A$ and $\partial B$ on which the disorder surface operators $\MW(\Sigma)$ are defined. Namely, its leading contributions should be given by the exchange of pairs of the lightest supergravity particles (smaller scaling dimensions $\Delta_k$), while its coefficients arise as a byproduct of the OPE coefficients appearing in the correlators of these surface operators with the chiral primary operators of the theory (see Figure 1). This proposal thus resembles the picture provided in \cite{maldacena}. 

\begin{figure}[t]
\centering
\includegraphics[width=11.5cm]{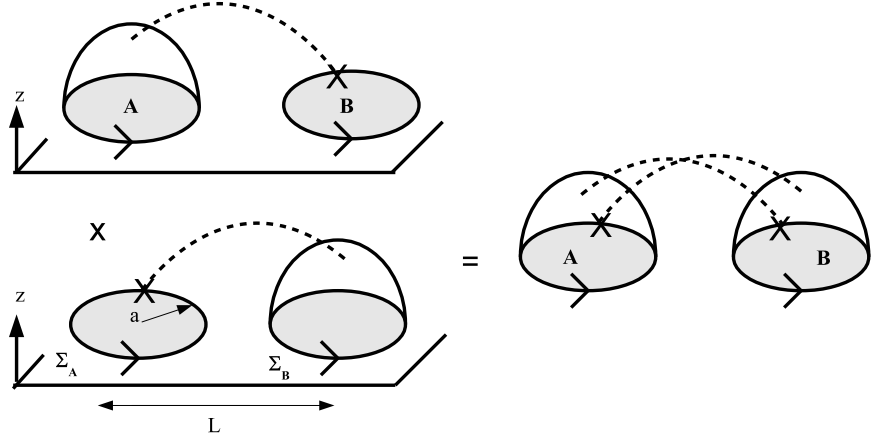}
\caption{Two static spherical 3-dimensional regions $A$ and $B$ (shaded grey) of radius $a$ separated by a long distance $L\gg a$ whose boundaries $\partial A$ and $\partial B$ define the surfaces $\Sigma_A$ and $\Sigma_B$ respectively (The figure is represented in one lower dimension for convenience). $z$ represents the \emph{radial} coordinate in AdS. Top Left: The emission of a supergravity particle (dotted line) from the D3-brane realization of the surface operator $\mathcal{W}(\Sigma_A)$ onto a point (X $\in \Sigma_B$) on the boundary of AdS where the CPO $\MO_B$ is inserted. Bottom Left: The emission of a particle from the D3-brane realization of the surface operator $\mathcal{W}(\Sigma_B)$ onto a point (X $\in \Sigma_A$) on the boundary of AdS where the CPO $\MO_A$ is inserted. Right: A leading contribution to the long distance OPE for $\mathcal{I}_{AB}$ is given by the exchange of a pair of the lightest supergravity particles between the surfaces $\Sigma_A$ and $\Sigma_B$.} 
\label{ope_ansatz}
\end{figure}

\subsection{Correlators of surface observables with local operators in the probe approximation}
We outline the procedure to compute the correlation function (\ref{eq9}). The coupling of the supergravity mode $s^{\Delta}$ (dual to $\MO_{\Delta}$) to the D3-probe brane realizing the operator $\MW(\Sigma)$ is given by a vertex operator $V_{\Delta}$. This can be determined by expanding the D3-brane action $S_{D3}=S_{D3}^{DBI} + S_{D3}^{WZ}$ to linear order in the fluctuations \cite{surface}. When the local operator $\MO_{\Delta}(\vec{x}\, ')$ emits the supergravity field $s^{\Delta}$ at a point $\vec{x}\, '$ on the boundary, if it contributes to the correlator of $\MO_{\Delta}$ with the surface operator $\MW(\Sigma)$, this supergravity mode propagates on the background AdS$_5 \times S^5$ and is then absorbed by the vertex operator, which must be integrated over the D3-brane realizing the operator $\MW(\Sigma)$ in AdS. The bulk field $s^{\Delta}$ has a simple propagator, however, it has a rather complicated set of couplings with the  supergravity fields accounting for the brane fluctuations. 

 In order to proceed, one may first write the scalar $s^{\Delta}$ in terms of a source $s_0^{\Delta}$ located at point $\vec{x}\, '$ on the boundary,
\beq
\label{eap1}
s^{\Delta}(\vec{x},z)=\int d^4\vec{x}\, '\, G_{\Delta}(\vec{x}\, ';\vec{x},z)\, s_0^{\Delta}(\vec{x}\, ').
\eeq
 Here, $G_{\Delta}(\vec{x}\, ';\vec{x},z)$ is the bulk to boundary propagator describing the propagation of the supergravity mode from the insertion point $\vec{x}\, '$ of the CPO to the point $(\vec{x},z)$ on the D3 probe brane,
\beq
\label{eap2} 
G_{\Delta}(\vec{x}\, ';\vec{x},z) = c\, \left(\frac{z}{z^2 + \vert \vec{x}-\vec{x}\, ' \vert^2}\right)^{\Delta},
\eeq 
 where the constant $c$ is fixed so as to require the normalization of the two-point correlation function $\langle \MO_{\Delta}\, \MO_{\Delta}\rangle$. As the surface operator $\mathcal{W}(\Sigma)$ is probed from a distance $L$ larger than its radius $a$, it is possible to approximate,
 \beq
 \label{eap3}
 G_{\Delta}(\vec{x}\, ';\vec{x},z) \simeq c\, \frac{z^{\Delta}}{ L^{2\Delta}}.
 \eeq 
 Then, it is necessary to write the fluctuations of $S_{D3}$ in terms of the field $s^{\Delta}$ given by Eq.(\ref{eap1}). This inmediatly leads to determine $V_{\Delta}$. Furthermore, it also allows to write the linearized fluctuation contribution of the D3-brane action as,
\beq
\label{eap4}
S_{D3} = T_{D3}\, \int d\MA\, V_{\Delta}\, s^{\Delta},
\eeq  
with $s^{\Delta}$ given in (\ref{eap1}) and $T_{D3}=N/2\pi^2$. In the last expression $d\MA$  refers to the volume element of the probe D3-brane. The correlation function is obtained from functionally differentiating the previous expression with respect to the source $s_0^{\Delta}$,
\barray
\label{eap5}
 \frac{\left\langle \mathcal{W}(\Sigma)\, \MO_{\Delta}(\vec{x}_0)\right\rangle }{\left\langle \MW(\Sigma) \right\rangle } & =& - \frac{\delta }{\delta s_0^{\Delta}(\vec{x}_0)} T_{D3}\int  d\MA\, d^4\vec{x}\, '\, V_{\Delta}\,  G_{\Delta}(\vec{x}\, ';\vec{x},z)\, s_0^{\Delta}(\vec{x}\, ') \\ \nonumber
 &=& -T_{D3}\, \int  d\MA\, V_{\Delta}\,  G_{\Delta}(\vec{x}_0;\vec{x},z).
\earray
If we let $\vec{x}_0$ to be parametrized as $(d_1 e^{i\phi_1}, d_2 e^{i\phi_2})$, then, integrating out this expression and using the approximation (\ref{eap3}) one thus obtains $\widetilde{\MC}_{\Delta}$ explicitly as \cite{yamaguchi},
\beq
\widetilde{\MC}_{\Delta, p}=\frac{2^{\Delta/2}}{\sqrt{\Delta}} C_{\Delta, p}\, \frac{(2 \pi \beta)^{\Delta}}{\lambda^{\Delta/2}}\, \frac{e^{-ip(\phi_1 + \phi_2)/2}}{(d_1 d_2)^{\Delta/2}}\, (1 + (-1)^{\Delta}),
\label{coeff_explicit}
\eeq
where $p=-\Delta, -\Delta+2, \cdots, 0, \cdots, \Delta$ is the momentum of the scalar field in $S^5$, $\beta$ is a parameter of the surface operator related with the geometric embedding of the D3-brane and $C_{\Delta,p}$ is a constant related with the spherical harmonics in $S^5$.

\subsection{Contributions from the lightest bulk fields}
For 10-dimensional supegravity compactified on AdS$_5 \times S^5$, the ten-dimensional fields may be written as,
\beq\label{SUGRA:fields}
\Psi=\sum_{p,I}\, \phi_p\, Y_{(p,I)},
\eeq
where $\phi_p$ is a five dimensional field and $Y_{(p,I)}$ are the spherical harmonics on $S^5$ with total angular momentum $p$. The full spectrum of 10D-supergravity compactified on $S^5$ was  obtained in \cite{sugra} but, in what follows, we will focus only in the lightest scalars $s^{\Delta}$, whose exchange will dominate the long distance behaviour of $\MI_{AB}$. These light scalar fluctuations couple to the $\MN=4$ SYM operators $\MO_{\Delta}$ of the lowest dimensions $\Delta$ which appear in the OPE for the surface operators $\MW(\Sigma)$ and $\MI_{AB}$. These states solve the Klein-Gordon equation in AdS$_5$,
\beq
\label{eq13}
\nabla_{\mu}\nabla^{\mu}s^{\Delta}=\Delta(\Delta-4)s^{\Delta}\quad \Delta\geq 2.
\eeq

 Note that the field $s^{\Delta}$ has a negative mass for $\Delta = 2,3$. However, these modes are not tachyonic, since they propagate on a space of negative curvature. In \cite{surface,yamaguchi} it has been shown in full detail how to obtain the correlator between a surface operator and the lightest of these fields, i.e the scalar with $\Delta=2$. For $p=0$ this yields,
\beq
\label{eq14}
\frac{\left\langle \mathcal{W}(\Sigma,0)\, \MO_{2,0}(L)\right\rangle }{\left\langle \mathcal{W}(\Sigma,0) \right\rangle } = \widetilde{\MC}_{2,0}\left( \frac{1}{L}\right)^{4}=\frac{1}{\sqrt{2}}\frac{(4\pi \beta)^2}{d_1 d_2}\frac{C_{2,0}}{\lambda}\left( \frac{1}{L}\right)^{4}
\eeq  
 which is of order $N^0$. From Eq.(\ref{eq14}) one may determine the contribution of the lightest scalar ($\Delta=2, p=0$) to $\MI_{AB}$.  This amounts to the leading contribution to the long distance expansion given in Eq.(\ref{su_mi:ope}). Defining $\kappa = \frac{C_{2,0}}{\sqrt{2}}\frac{(4\pi \beta)^2}{d_1 d_2}$, this expansion reads as,
 \beq
 \label{eq16}
 \MI_{AB}\sim  \left(\widetilde{\MC}_{2}\right)^2\, \left( \frac{1}{L}\right)^{8}\, + \cdots\,  = \frac{\kappa^2}{\lambda^2}\, \left( \frac{1}{L}\right)^{8}\, +\,  \MO(L^{-4\Delta}, \Delta \geq 3),
 \eeq
 which only depends on $\lambda$. 
 
 As a result, it has been checked that the leading order of the long distance $\MI_{AB}$ provided by the OPE (\ref{su_mi:ope}), is $ (G_N^{\, (5)})^{\, 0}\sim N^{\, 0}$. This $N$ dependence is subleading with respect to the expected $N^{\, 2}$ dependence which  holds when a fully connected minimal surface $\gamma_{A \cup B}^{con}$ between the regions $A$ and $B$ is allowed in an holographic computation. Thus, the holographic mutual information $\MI_{AB}$ experiences a phase transition marked by a change in the $N$ dependence of its leading contributions but does not suffer a sharp vanishing due to large $N$ effects. Namely, it smoothly decays following a power law given by Eq.(\ref{eq16}) while parametrically saturates the bound given by Eq.(\ref{eq_wolf}). 

\section{Conclusions}
In this note, we have investigated the structure of the quantum corrections to the holographic mutual information $\MI_{AB}$ between two wide separated regions in the $\MN=4$ SYM gauge theory dual to AdS$_5 \times S^5$. To this end, first we have recasted the correlators of twist field operators related to the computation of the mutual information, in terms of correlators between \emph{surface} operators in gauge theories. Namely, it is reasobable enough to claim that the twist field operators in a $d+1$ theory would be some kind of codimension-2 disorder-like surface operators. As so little is known about the higher dimensional versions of the twist field operators, here we have only relied on the most basic analogies between them and the disorder-like surface operators.  It is worth to note that, by no means we have tried to establish an exact identification between them. Further investigations in this direction are surely needed in order to obtain some explicit (holographic or field theoretical) constructions of the twist operators in higher dimensions. In spite of this, we feel that the commented analogies are strong enough to obtain valuable information about the $N$-dependence of the first non vanishing quantum corrections to the mutual information.  Under this assumption, we have used the AdS/CFT realizations of the surface operators in the probe approximation, to provide a long distance expansion for the $\MI_{AB}$. The coefficients of this expansion arise as a byproduct of the OPE for the correlators of the surface operators with the chiral primary operators of the theory. The results show that in the case under consideration, the mutual information $\MI_{AB}$ undergoes a phase transition at a critical distance marked by a change in the $N$ dependence of its leading contributions. Namely, in the large separation regime $\MI_{AB}\sim\MO(N^0)$, so instead of strictly vanishing, it smoothly decays with a power law shaped by the exchange of pairs of the lightest bulk particles between $A$ and $B$.

\section*{Acknowledgments}
The author is grateful to G. Sierra, A.V. Ramallo and E. Tonni for giving very 
valuable insights at different stages of this project. JMV has been 
supported by Ministerio de Econom\'ia y Competitividad Project 
No. FIS2012-30625. I thank A.V. Ramallo and J. Mas 
for their hospitality at Universidad de Santiago de Compostela where 
this project took its initial steps.

\end{document}